\documentclass[aps,prb,twocolumn,superscriptaddress,superscriptaddress]{revtex4-2} 

\usepackage{graphicx}  	% needed for figures
\usepackage{dcolumn}   	% needed for some tables
\usepackage{bm}        	% for math
\usepackage{amssymb}   	% for math
\usepackage[dvipsnames]{xcolor}	   % for colored text
\usepackage{siunitx}   	% for unit definitions
\usepackage{physics}  	% for equations

\usepackage{multibib}
\newcites{supp}{Supplementary References}

\begin{document}
%\rightline{\today}
\title{Spectroscopic signature of surface states and bunching of bulk subbands in topological insulator (Bi$_{0.4}$Sb$_{0.6}$)$_2$Te$_3$ thin films}

\author{L.~Mulder}
\thanks{These authors contributed equally to this work.}
\affiliation{MESA+ Institute for Nanotechnology, University of Twente, The Netherlands}
\author{C.~Castenmiller}
\thanks{These authors contributed equally to this work.}
\affiliation{MESA+ Institute for Nanotechnology, University of Twente, The Netherlands}
\author{F.J.~Witmans}
\affiliation{MESA+ Institute for Nanotechnology, University of Twente, The Netherlands}
\author{S.~Smit}
\affiliation{Van der Waals-Zeeman Institute, Institute of Physics, University of Amsterdam, The Netherlands}
\author{M.S.~Golden}
\affiliation{Van der Waals-Zeeman Institute, Institute of Physics, University of Amsterdam, The Netherlands}
\author{H.J.W.~Zandvliet}
\affiliation{MESA+ Institute for Nanotechnology, University of Twente, The Netherlands}
\author{P.L.~de Boeij}
\affiliation{MESA+ Institute for Nanotechnology, University of Twente, The Netherlands}
\author{A.~Brinkman}
\affiliation{MESA+ Institute for Nanotechnology, University of Twente, The Netherlands}
	
\begin{abstract}
High quality thin films of the topological insulator (Bi$_{0.4}$Sb$_{0.6}$)$_2$Te$_3$ have been deposited on SrTiO$_3$ (111) by molecular beam epitaxy. Their electronic structure was investigated by \textit{in situ} angle-resolved photoemission spectroscopy and \textit{in situ} scanning tunneling spectroscopy. The experimental results reveal striking similarities with relativistic \textit{ab-initio} tight binding calculations. We find that ultrathin slabs of the three-dimensional topological insulator (Bi$_{0.4}$Sb$_{0.6}$)$_2$Te$_3$ display topological surface states, surface states with large weight on the outermost Te atomic layer, and dispersive bulk energy levels that are quantized. We observe that the bandwidth of the bulk levels is strongly reduced. These bunched bulk states as well as the surface states give rise to strong peaks in the local density of states. 
\end{abstract}

\maketitle
\section{Introduction}
Chalcogenide three-dimensional (3D) topological insulators (TI), such as Bi$_2$Se$_3$, Bi$_2$Te$_3$ and Sb$_2$Te$_3$, are well-known for their topological surface states \cite{H.Noh2008EPL,Y.Xia2009NP,Y.Chen2009S}. These metallic surface states, which are characterized by a linear dispersion as well as spin-momentum locking, appear as a Dirac cone at the $\Gamma$-point. Since Bi$_2$Te$_3$ single crystals are intrinsically \textit{n}-doped, and Sb$_2$Te$_3$ \textit{p}-doped, alloys of the two have been developed in order to engineer a material in which the chemical potential is located within, or close to, the bulk band gap \cite{X.He2012APL}. An advantage of a higher Sb/Bi ratio, is the shift of the Dirac point (DP) to higher energies, reaching above the top of the bulk valence band (VB) \cite{J.Zhang2011NC}.\\\\
Due to the spatial extent of the surface state in the \textit{z}-direction, it has been predicted that ultrathin slabs of 3D TIs provide a route to a hybridization gap around the DP \cite{J.Linder2009PRB,C.Liu2010PRB81}, thereby gapping out the topological surface states and paving the way to topologically protected one-dimensional (1D) edge state transport, very comparable to the quantum spin Hall (QSH) effect in HgTe quantum wells \cite{B.Bernevig2006S,M.Konig2007S,B.Zhou2008RPL}. Even though Bi$_2$Se$_3$ presents the largest bulk bandgap, Bi$_2$Te$_3$ and Sb$_2$Te$_3$ are predicted to be more appealing candidates to exhibit 1D transport due to their relatively short edge state decay length in the ultrathin film limit \cite{C.Liu2010PRB81,M.Kim2012PNAS,M.Asmar2018PRB}. While stoichiometric slabs have been well-investigated spectroscopically \cite{Y.Jiang2012PRL1,L.Plucinski2013JAP,P.Ngabonziza2015PRB}, even in the limit of hybridized surface states \cite{Y.Jiang2012PRL6}, the electronic structure of ultrathin films of off-stoichiometric (Bi$_{1-x}$Sb$_{x}$)$_2$Te$_3$ is still relatively unexplored. Zhang \textit{et al.} \cite{J.Zhang2011NC} studied the influence of the stoichiometry on the electronic properties of ultrathin (Bi$_{1-x}$Sb$_{x}$)$_2$Te$_3$ by means of angle-resolved photoemission spectroscopy (ARPES) and transport measurements. They show that band structure can be engineered by changing the stoichiometry. Furthermore, quasiparticle interference patterns have been imaged with a scanning tunneling microscope, revealing the dispersion above the Fermi energy ($E_F$) \cite{X.He2015SR,K.Scipioni2018PRB}.\\\\
Here, we investigate the electronic structure of few-nm (Bi$_{0.4}$Sb$_{0.6}$)$_2$Te$_3$ by directly comparing the results of multiple techniques. We performed ARPES measurements, scanning tunneling spectroscopy (STS) measurements and \textit{ab-inito} tight-binding (TB) calculations. The good agreement between the measured and calculated electronic structure reveals a bunching effect of the bulk sub-levels, the identification of Bi/Sb topological surface states, as well as Te surface states close to the top of the VB, leading to strong peaks in the density of states (DOS). The modeling shows that for (Bi$_{1-x}$Sb$_{x}$)$_2$Te$_3$ with $x$ = 0.6, the reduced band width is likely caused by a Sb/Bi substitution-induced inversion of the band order near the high-symmetry point $\Gamma$. This inversion interchanges the bands close to $E_F$, while the order at the Z-point remains unaltered.

\section{Materials and methods}
\textbf{Molecular beam epitaxy\quad} For this study, 5 and 10 \si{nm} (Bi$_{0.4}$Sb$_{0.6}$)$_2$Te$_3$ films were deposited using molecular beam epitaxy (MBE) on gate-tunable, Ti-terminated SrTiO$_3$ (111) substrates in a deposition chamber with a base pressure of about $5.0\times 10^{-11}$ \si{\milli\bar}. Details on the substrate treatment \cite{G.Koster1998APL} can be found in the Supplemental Material (SM) \cite{SM}. High-purity Bi (6N), Sb (6N) and Te (6N) were evaporated from standard Knudsen effusion cells. During the deposition, the substrate was held at a temperature of 225 \si{\degreeCelsius}. A quartz crystal microbalance was used to calibrate the individual material fluxes. To ensure a high crystalline growth while suppressing the amount of Te vacancies, we employed a flux ratio of (Bi+Sb):Te = 1:10, with a deposition rate of 0.07 \si{\nm}/\si{\minute}. To verify the stoichiometry of the deposited film, \textit{in situ} X-ray photoelectron spectroscopy (XPS) spectra were recorded using an Omicron nanotechnology surface analysis system, equipped with a monochromatic aluminium source (K$\alpha$ X-ray source XM1000). \\\\
\textbf{Angle-resolved photoemission spectroscopy \quad} ARPES data were acquired at the Van der Waals-Zeeman Institute. The ARPES spectra are taken at a base temperature of 15 \si{\kelvin}, using an L1 He-Lamp, emitting linearly, \textit{p}-polarized photons with an energy of 21.2 \si{\electronvolt}, and an A-1 hemispherical analyser, both from MB-Scientific. The system is equipped with a TMM 304 UV monochromator from Specs. The energy resolution was 15 \si{\milli\electronvolt} and the angular resolution better than 0.1\si{\degree}. Laser-ARPES spectra were obtained using a fourth harmonic source from APE GmbH, producing 200 \si{\nm} photons with a corresponding energy of 6.2 \si{\electronvolt}, with an energy resolution of 5 \si{\milli\electronvolt} and a k-resolution of 0.002 \si{\per\angstrom}. The measurements were performed while maintaining a pressure of $<5.0\times 10^{-11}$ \si{\milli\bar}.\\\\
\textbf{Scanning tunneling spectroscopy\quad} Scanning tunneling microscopy (STM) and spectroscopy measurements were performed at 77 \si{\kelvin} with a low-temperature Omicron ultra-high vacuum (UHV) STM. The base pressure of the STM chamber was below $1.0\times 10^{-11}$ \si{\milli\bar} and PtIr tips were used. The \textit{I(V)} spectroscopy experiments were conducted in the constant height mode, i.e. the tunnel current was measured while the sample bias was ramped with the feedback loop disabled. The (normalized) \textit{dI(V)/dV} curves are the numerical derivatives of the \textit{I(V)} recordings. The \textit{dI(V)/dV} and \textit{(dI(V)/dV)/(I(V)/V)} curves were obtained by averaging over many \textit{I(V)} traces recorded at different locations.\\\\
\textbf{\textit{Ab-initio} tight binding modeling \quad} The electronic structure of the ternary tetradymite alloy was obtained using a TB approach based on the \textit{ab-initio} Greens-function calculations of Aguilera \textit{et al.} \cite{I.Aguilera2019PRB}. Their projected TB parameters turn out to be very similar for pristine Bi$_2$Te$_3$ and Sb$_2$Te$_3$ \cite{footnote}, resulting in very comparable band dispersions, apart from band inversions along the line $\Gamma-$Z. This close resemblance allows us to interpolate to the alloy system in the following way: the quasiparticle states for the alloy are expanded as wave packets of Bloch-like sums of localized functions, 
\begin{equation}
\psi(\mathbf{r}) = \sum_{m\mathbf{k}} c_{m\mathbf{k}}\phi_{m\mathbf{k}}(\mathbf{r}),
\end{equation}
\begin{equation}
\phi_{m\mathbf{k}}(\mathbf{r}) = \frac1{\sqrt{N}}\sum_\mathbf{R} e^{i\mathbf{k}\cdot\mathbf{R}} w_m(\mathbf{r}-\mathbf{R}).
\end{equation}
The $w_m(\mathbf{r})$ centred on a Bi- or on a Sb-atom are derived from respectively the projected Wannier functions of the pristine Bi$_2$Te$_3$ or Sb$_2$Te$_3$ crystals, and when on a Te-atom, from either one. The alloy disorder is thus incorporated into this mixed basis, which can reasonably be assumed to inherit the orthogonality of the pristine Wannier functions. The alloy Hamiltonian elements on the $\mathbf{k}$-diagonal then reduce to the weighted average of the pristine ones. As these are by far the largest terms, they effectively define a virtual crystal (VC) Hamiltonian for the alloy. The alloy disorder further enters via smaller off-$\mathbf{k}$-diagonal elements that may be accounted for in a perturbative expansion. These off-diagonal terms will cause the wave packets to spread over a range of $\mathbf{k}$-vectors centred around the VC one. As turns out, it suffices to compute the VC-band structure, $\{\epsilon_{n\mathbf{k}}, \psi_{n\mathbf{k}}\}$, to identify the main features in the ARPES measurements. Similarly we compute the laterally averaged local density of states (LDOS) from the VC-dispersions, and use the Tersoff-Hamann model \cite{J.Tersoff1985PRB} to simulate the marginal tunnel currents,
\begin{equation}
\frac{d I}{dV} \propto \sum_{n\mathbf{k}} \delta(\epsilon_{n\mathbf{k}}-eV)e^{-2\kappa_{n\mathbf{k}}z}|\langle\psi_{n\mathbf{k}}\mid \mathrm{Te} \rangle|^2.\label{tunnel_currents}
\end{equation}
Here the quasiparticle amplitudes at the outermost Te-atoms are used to generate the LDOS at the tip position $z$ (9\AA), while accounting for the $\mathbf{k}$- and $\epsilon_{n\mathbf{k}}$-dependent inverse decay length $\kappa_{n\mathbf{k}}$, satisfying
\begin{equation}
\kappa_{n\mathbf{k}} = \sqrt{\frac{2m}{\hbar^2} \big(\Phi-\frac12\epsilon_{n\mathbf{k}}\big) + k_\parallel^2},\label{eq:kappa}
\end{equation}
in which $\Phi$ (5.4 \si{\electronvolt}) is the work function. This simulated signal can be decomposed into layer contributions, as is done in Fig. \ref{fig:STS}(b), by adding a projective weight $\nu_{n\mathbf{k}} = |\langle\psi_{n\mathbf{k}}\mid \mathrm{A}\rangle|^2$ to the summand for atomic layer A. The same projective weight is used to identify and color-code the surface states in the band dispersion in Figs. \ref{fig:Stoichiometry}, \ref{fig:ARPES}(b), \ref{fig:STS}(a), and \ref{fig:Thickness-SM}.\\\\
\textbf{Vacuum transfer\quad } To avoid surface contamination and oxidation we have transferred the thin films from  the MBE system to the XPS, STM and ARPES systems using a UHV suitcase. The vacuum suitcase is equipped with a non-evaporable getter pump and an ion pump, which together ensured the pressure remained below $1.5\times 10^{-10}$ \si{\milli\bar} during the transfer process. In the ARPES system, the film has briefly been exposed to a pressure of $3.0\times 10^{-10}$ \si{\milli\bar} during the transfer of the sample from the suitcase to the main chamber. In order for a film to arrive at the STM main chamber the sample has to move through a load lock and preparation chamber with a base pressure of $5.0\times 10^{-9}$ \si{\milli\bar} and $1.0\times 10^{-10}$ \si{\milli\bar}, respectively. 

\section{Results and discussion}

\begin{figure*}
\centering
\includegraphics[width=\textwidth]{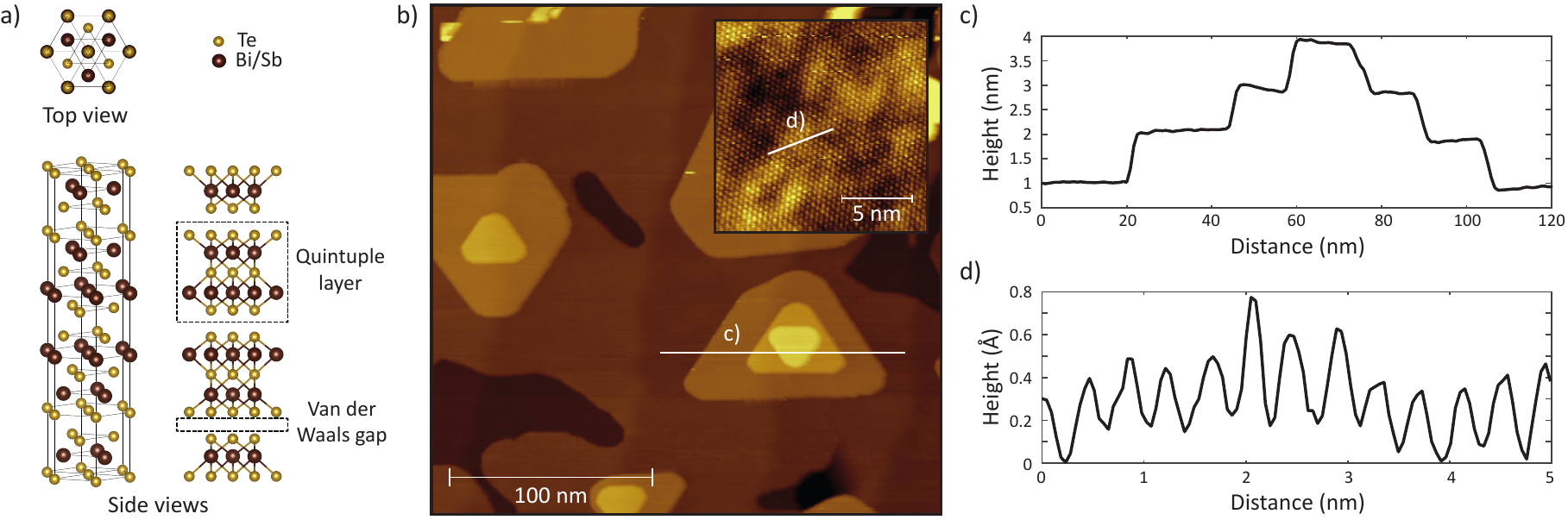}
\caption{\label{fig:Topography} (a) The layered rhombohedral tetradymite crystal structure of (Bi$_{1-x}$Sb$_{x}$)$_2$Te$_3$ revealing the quintuple layer stacking along the c-axis, which are weakly bonded via Van der Waals interaction. (b) STM topography image (250 x 250 \si{nm}) of a 5 \si{nm} (Bi$_{0.4}$Sb$_{0.6}$)$_2$Te$_3$ film measured at a bias voltage of 1.0 \si{\volt} and a current setpoint of 400 \si{\pico\ampere}. The inset shows a STM topography image (15 x 15 nm) with the atomic structure on a single terrace measured at a bias voltage of 0.2 \si{\volt} and a current setpoint of 1.0 \si{\nano\ampere}. The hexagonal structure as shown in panel (a) is clearly visible. (c) Line profile taken at the white line of the main figure of panel (b) which crosses a stack of triangular terraces. The height difference at every step edge is approximately 1 \si{nm}. The middle terrace is slightly tilted since it is located at a 0.2 \si{nm} high substrate terrace step edge. (d) Line profile taken at the white line of the inset of panel (b).}
\end{figure*} 
(Bi$_{1-x}$Sb$_{x}$)$_2$Te$_3$ crystallizes in the rhombohedral tetradymite crystal structure, as depicted in Fig. \ref{fig:Topography}(a). The (Bi$_{1-x}$Sb$_{x}$)$_2$Te$_3$ surface-matched unit cell is composed of three quintuple layers (QLs) each of which are approximately 1 \si{nm} high and separated from each other by a Van der Waals gap. The QLs consist of five atomic layers in a Te-X-Te-X-Te stacking, where X can either be Bi or Sb. The STM constant current image in Fig. \ref{fig:Topography}(b) reveals the film's surface morphology. The film exhibits triangular islands, reflecting the three-fold symmetry of the crystal structure, with single QL terrace step edges, visible in the line profile shown in Fig. \ref{fig:Topography}(c). The 0.2 \si{nm} height differences observed in the background arise from individual Ti-terminated SrTiO$_3$ (111) terrace step edges. The inset of Fig. \ref{fig:Topography}(b) shows a constant current STM image with atomic resolution. The hexagonal atomic arrangement matches the materials crystal structure. Additionally, the extracted line profile, presented in Fig. \ref{fig:Topography}(d), reveals an interatomic distance which is in good agreement with those reported for Bi$_2$Te$_3$ and Sb$_2$Te$_3$, which exhibit a lattice parameter of approximately 4.39 \si{\angstrom} \cite{S.Nakajima1963JPCS} and 4.26 \si{\angstrom} \cite{T.Anderson1974ACSB}, respectively. The observed global height inhomogeneity is allocated to disorder induced charge density fluctuations \cite{X.He2015SR,K.Scipioni2018PRB,B.Jack2021PRR}.

\subsection*{\textit{Ab-initio} tight binding modeling}

\begin{figure*}
\centering
\includegraphics[width=\textwidth]{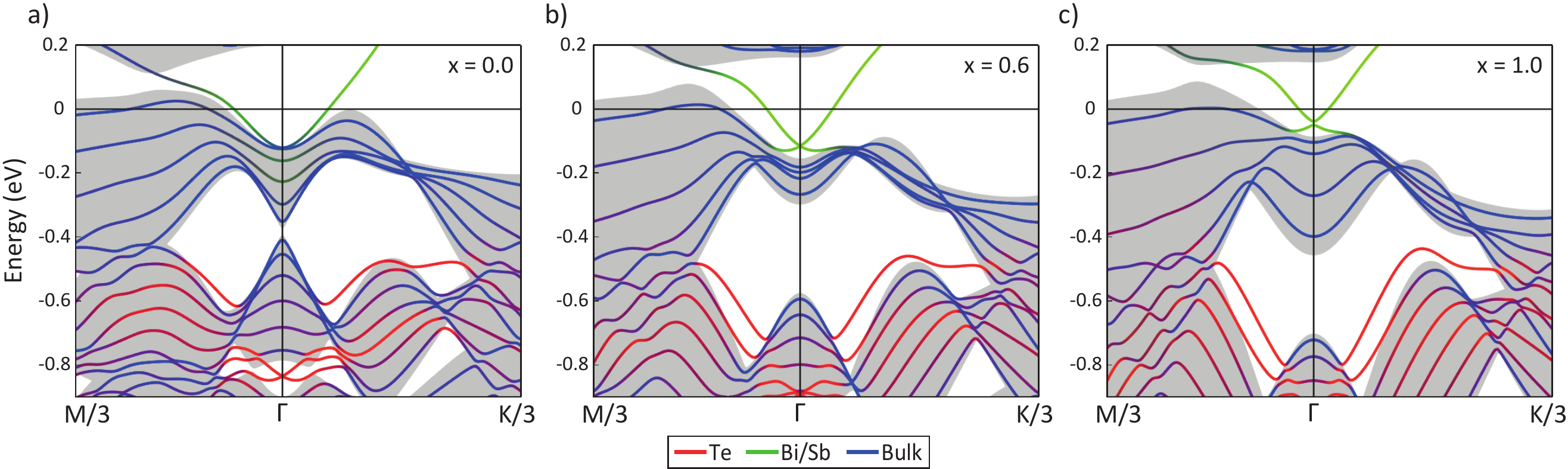}
\caption{\label{fig:Stoichiometry} \textit{Ab-initio} TB calculations for a 6 QL free-standing thin slab of (Bi$_{1-x}$Sb$_{x}$)$_2$Te$_3$ in the $\text{M}-\Gamma-\text{K}$ direction, for (a) $x$ = 0.0, (b) $x$ = 0.6 and (c) $x$ = 1.0. The energy is plotted with respect to the predicted $E_F$. The green, red and blue lines respectively represent Bi/Sb surface states, Te surface states and states with a bulk character. The grey area marks the projected bulk bands. All states outside this projection represent surface states. The calculations visualize the characteristic shift of the DP with respect to the top of the VB. Furthermore, they show that for mixed stoichiometries, especially close to $x$ = 0.6, the bulk states bunch at an energy close to -0.2 \si{\electronvolt}.}
\end{figure*}

The electronic structure of the (Bi$_{1-x}$Sb$_x$)$_2$Te$_3$ alloys was obtained within the VC approximation of the \textit{ab-initio} TB model as outlined above. All slab calculations were performed on free-standing vacuum-surrounded slabs. For the modeling we take into account all six $p$-type spin orbitals of every atom in the unit cell, which contains only a single QL for the 3D bulk systems. The covalent bonding inside the QL gives rise to 15 doubly degenerate VBs, of which 9 are occupied and 6 unoccupied, separated by a small gap along the line F-Z. We refer to Zhang \textit{et al.} for a schematic of both the bulk and surface Brillouin zone, showing labels for the high symmetry points \cite{H.Zhang2009NP}. The dispersion along the line $\Gamma$-Z, i.e. in the direction perpendicular to the QLs, is small due to the van der Waals bonding type in between the QLs. In the pristine Bi$_2$Te$_3$ bulk crystal we find that the top VB has an energy at $\Gamma$ that is well below its energy at the Z-point. In the pristine Sb$_2$Te$_3$ the reverse order is obtained. Upon sufficient alloying, we find that in our VC calculations the band order along $\Gamma$-Z can be tuned, causing the band dispersion along this line to nearly vanish at intermediate alloy fractions close to 0.6. This has an important effect on the electronic structure of the 2D slabs containing small numbers of QLs. In such systems, the small thickness causes the k-vector to quantize along $\Gamma$-Z, and the 2D-projected bulk bands to discretize proportional to the number of QLs, with an energy separation that depends on the alloy fraction $x$. A flat bulk dispersion along $\Gamma$-Z, causes the bulk-like bands to bunch, as shown in Fig. \ref{fig:Stoichiometry}, with a striking reduction of the bandwidth around the $\Gamma$-point visible in panel (b).\\\\
In addition, topological surface states develop from the bulk bands, which separate from these bulk-like bands into the gaps that open near $\Gamma$. These surface states are only weakly sensitive to the number of QLs. As these surface states live on either side of the slab, and decay exponentially into the bulk, the development of the bonding - anti-bonding gap at $\Gamma$ is indicative for the decay length. We estimate the decay length to be about 3 QLs, with an odd-even difference due to the nodal plane inside or in between the middle QLs.\\\\
The electronic structure of the slabs, as depicted in Figs. \ref{fig:Stoichiometry}, \ref{fig:ARPES}(b), \ref{fig:STS}(a) and \ref{fig:Thickness-SM} are indeed mainly composed of closely spaced parallel bands inside each projected 3D-bulk band (grey) with predominantly bulk character (blue), and a fixed number of surface states, that lie outside the projected bulk bands. These surface states are colored green if they have a large weight on the outermost Bi/Sb atoms, and red if on the outermost Te layer. The Bi/Sb surface state develops a Dirac cone at $\Gamma$ for increasing slab thicknesses, as shown in Fig. \ref{fig:Thickness-SM}, and will be termed topological surface state (TSS). The reversal of the band order at $\Gamma$ causes the DP to shift from buried inside the VB for Bi$_2$Te$_3$ to slightly above the VB for Sb$_2$Te$_3$, see Fig. \ref{fig:Stoichiometry}.\\\\
In the VC approximation the top-valence bands have small energy dispersions over a considerable portion of the Brillouin zone. Such nearly flat bands can easily be localized, and will thus be more sensitive to the local disorder in the alloy. This disorder needs to be included in the perturbation expansion going beyond the VC approximation. Furthermore, the additional degeneracy caused by the bunching of these flat, bulk-like bands brings about localisation not just in the lateral direction, but also in the perpendicular direction. This makes them even more susceptible to the alloy disorder. The vicinity of these bands to the DP may cause hybridisation of these bands with the TSS, potentially interfering with the delocalized nature of the TSS.

\subsection*{Angle-resolved photoemission spectroscopy}
\begin{figure*}
\centering
\includegraphics[width=0.65\textwidth]{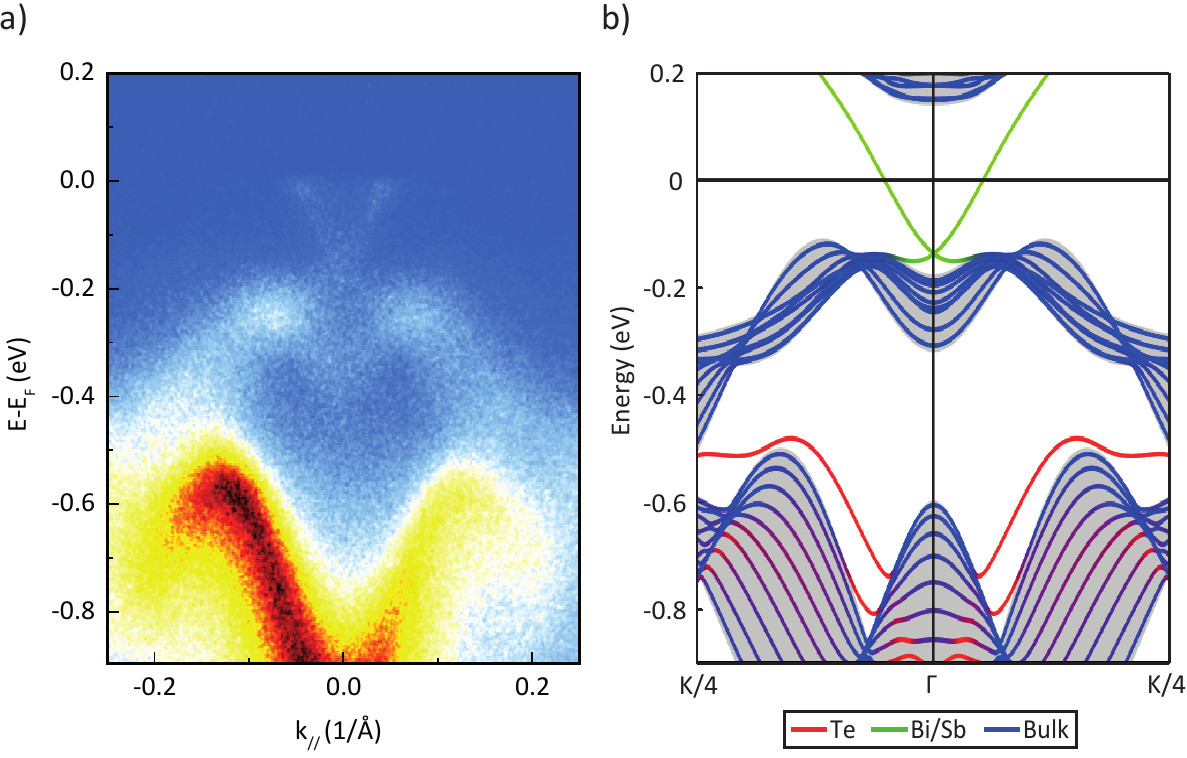}
\caption{\label{fig:ARPES} Correlation between TB calculations and ARPES measurements (a) ARPES EDM of the band dispersion along the $\bar{\text{K}}-\bar{\Gamma}-\bar{\text{K}}$ direction of a 10 \si{nm} (Bi$_{0.4}$Sb$_{0.6}$)$_2$Te$_3$ film deposited on a SrTiO$_3$ (111) substrate. The V-shaped dispersion, at energies above $-200$ \si{\milli\electronvolt}, originates from the TSS. The relatively weak M-shaped dispersion just below it, at $-200$ \si{\milli\electronvolt} $< E-E_F < -400$ \si{\milli\electronvolt}, is from the bulk VB. At energies below $-500$ \si{\milli\electronvolt} the Te surface state shows up, which is flanked by bulk-like intensity at lower energies. The DP is located in vicinity of the top of the bulk VB. (b) \textit{Ab-initio} TB calculations for a free-standing thin slab, 10 QL of (Bi$_{0.4}$Sb$_{0.6}$)$_2$Te$_3$, in the $\text{K}-\Gamma-\text{K}$ direction. The y-axis presents the energy of the electronic states with respect to the predicted $E_F$, resulting from the TB calculations. To allow for a proper comparison between the experimentally observed and theoretically predicted band structure, the TB slab calculation is plotted up until a crystal momentum of K/4 which corresponds to the measured range of 0.25 \si{\per\angstrom} in the ARPES EDM.}
\end{figure*}

To study the electronic band structure, ARPES measurements have been performed on a nominal 10 \si{nm} (Bi$_{0.4}$Sb$_{0.6}$)$_2$Te$_3$ film. The resulting electron distribution map (EDM) measured along the $\bar{\text{K}}-\bar{\Gamma}-\bar{\text{K}}$ direction is presented in Fig. \ref{fig:ARPES}(a). On the same film, laser-ARPES measurements were conducted. These results confirm the linearly dispersing topological surface states, along with a circular constant energy contour throughout the occupied part of Dirac cone, see Fig. \ref{fig:ARPES-SM}. The conformity of the TB model calculations on a vacuum-surrounded slab to the ARPES data from a film of similar thickness, shown in Fig. \ref{fig:ARPES}, is striking. The small discrepancy between the $E_F$ in the ARPES EDM and the TB slab calculation can be explained by surface adsorption of residual gases, which is likely to take place when the sample is cooled down to 15 \si{\kelvin}, at which temperature the ARPES measurements were performed \cite{C.Chen2012PNAS}. Additionally, the effect can also be attributed to defects present in the (Bi$_{0.4}$Sb$_{0.6}$)$_2$Te$_3$ film. A more in-depth description regarding the observed discrepancy in $E_F$ will follow in the next section. The TSS are relatively weak, but clearly resolved in the ARPES EDM at energies above $-200$ \si{\milli\electronvolt}, and merge on the lower binding energy side with weak bulk-like states between $-200$ \si{\milli\electronvolt} and $-400$ \si{\milli\electronvolt}. The Te surface state shows up as a much more intense feature in the ARPES below $-500$ \si{\milli\electronvolt}, with an expected dispersion and flanked at the lower energy side by bulk-like intensity. The difference in intensity between the surface states and the bulk bands in the EDM can be explained by taking into consideration that, for a photon energy of 21.2 \si{\electronvolt}, the escape depth of the emitted electron is of the order of only a couple of nm. This means that we expect the surface states with large weight on the outermost Te atomic layer to be more intense than those with large weight on the Bi/Sb layer beneath. Moreover, we also expect these surface states to be more intense than the bulk-like states from the layers below, even though the latter are greater in number. In particular we observe that the Te surface state is asymmetrically sensitive to the $p$-polarized light, while the bulk like states and the TSS are more symmetric. A more detailed description is provided in the SM.\\

\noindent The location of the DP, $E_D$, with respect to the top of the VB, along $\text{K}-\Gamma-\text{K}$, as found in the TB calculations is slightly different than observed in ARPES. From the ARPES EDM we find the DP to lie approximately 50 \si{\milli\electronvolt} above the top of the VB, whereas in the TB slab calculation $E_D$ coincides with the top of the bunched VB bulk states in the $\Gamma-\text{K}$ direction. By extracting the Dirac fermion velocity from the linear dispersion around the DP from the ARPES EDM, using an undistorted upper cone fitted to the $\bar{\Gamma}-\bar{\text{K}}$ dispersion angle, we find $v_F$ to be about 4.7$\times 10^5$ \si{\metre/\second}. By applying the same procedure on the TB calculated TSS dispersion, we find a $v_F$ of 4.1$\times 10^5$ \si{\metre/\second}, which closely resembles the experimental value. Additionally, these values are also in good correspondence with previously published calculations \cite{H.Zhang2009NP} and ARPES data \cite{X.He2012APL}. However, in the TB slab calculations, we note that at energies just below $E_D$, the TSS present a flat-band behavior, causing them to exhibit a non-linear behavior at energies close to the VB.
\subsection*{Scanning tunneling spectroscopy}
\begin{figure*}
\centering
\includegraphics[width=\textwidth]{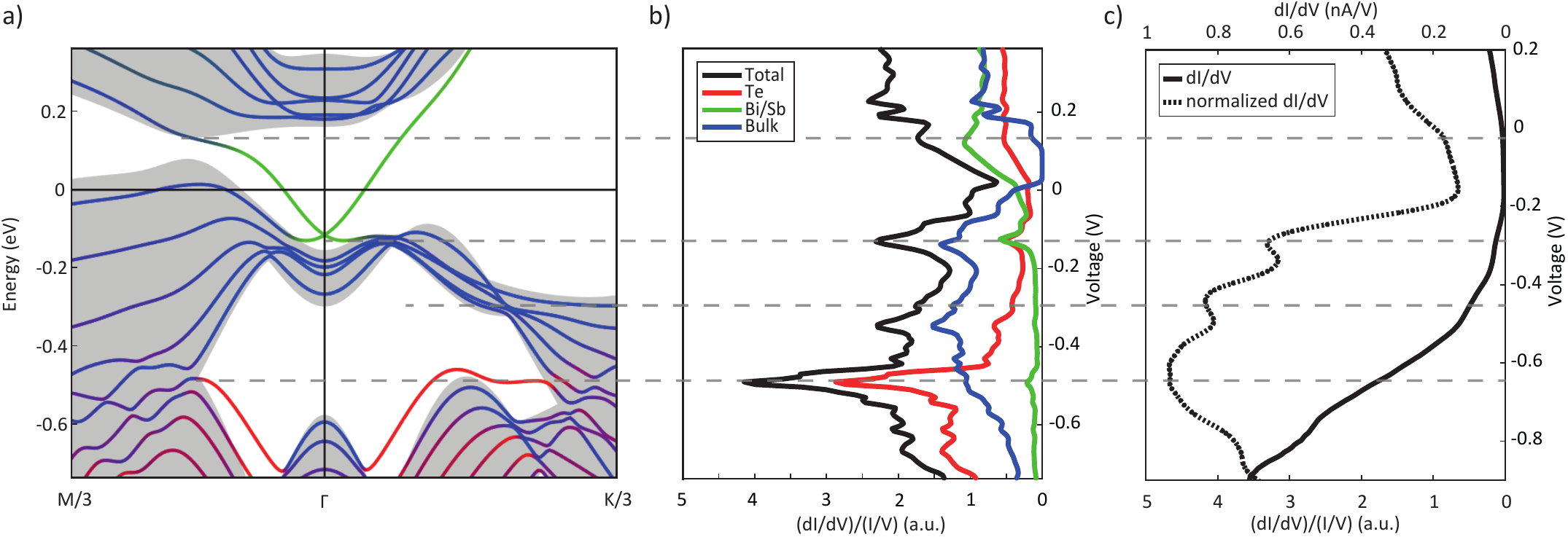}
\caption{\label{fig:STS} \textit{(dI/dV)(I/V)} spectra of a (Bi$_{0.4}$Sb$_{0.6}$)$_2$Te$_3$ film as a function of applied bias voltage. (b) STS spectrum based on TB calculations of a 6 QL thin slab (a), presenting the contribution of the individual atoms, the bulk and total contribution to the STS. (c) \textit{dI/dV} and corresponding DOS of a 5 \si{\nm} (Bi$_{0.4}$Sb$_{0.6}$)$_2$Te$_3$ film on SrTiO$_3$ as a function of applied bias voltage. In order to allow for a direct comparison between the theoretical predictions by TB calculations (a)-(b) and experimental STS data, the energy/bias voltage window of the TB results has been adapted to match the STS spectrum. The oscillations observed in the DOS coincide with both bunched bulk states, near -0.30 \si{\volt} and -0.43 \si{\volt}, as well as the Te contribution to the surface states, near -0.63 \si{\volt}. }
\end{figure*}

The sub-band spacing in Sb$_2$Te$_3$, determined by quantum confinement, was previously shown to give rise to weak oscillations in the DOS, as revealed by STS \cite{Y.Jiang2012PRL6}. The quantum well states were shown to vary with the thickness of the slab. We anticipate that the situation is very different for (Bi$_{1-x}$Sb$_{x}$)$_2$Te$_3$: the bunched nature of the quantum well states should give a strong signal in STS, and the peak positions are mainly given by the energies where the bands are flat. In (Bi$_{0.4}$Sb$_{0.6}$)$_2$Te$_3$, this bunching effect of flat bands turns out to occur, at comparable $E-E_F$, independent of the slab thickness, see Fig. \ref{fig:Thickness-SM}.\newpage%\\\\
\noindent STS data can, in some cases, also provide information on the dispersion of the energy bands \cite{Z.Jiao2021APL}, as elucidated in the SM. We show that the extracted Fermi velocity, $v_F$, from the voltage dependence of the inverse decay length in the vicinity of the DP gives a $v_F$ of about 1.5$\times 10^5$ \si{\metre/\second}. Note, that this value is significantly lower than the $v_F$ extracted from the ARPES EDM, because the latter $v_F$ was extracted higher up in the cone. This is in excellent agreement with the approximately 2.4$\times 10^5$ m/s found using our TB model by considering the apex to be at -0.15 \si{\electronvolt} relative to $E_F$, and the tangent at the DP.
\\\\
We use the Tersoff-Hamann model to simulate the tunnel currents using the LDOS derived from our TB model \cite{J.Tersoff1985PRB}. This way we are able to resolve the contribution of the electronic bands to the total dI/dV of an STS spectrum. The band structure for a free-standing 6 QL (Bi$_{0.4}$Sb$_{0.6}$)$_2$Te$_3$ film is presented in Fig. \ref{fig:STS}(a). The resulting simulation of the corresponding STS spectrum is shown in Fig. \ref{fig:STS}(b). The k$_\parallel$-dependent decay length of the states into the vacuum will cause the STS to be most sensitive to the dispersions close to the $\Gamma$ point, and in particular to the ones giving rise to van Hove singularities. This is clearly observed in the computed normalized dI/dV curves in Fig. \ref{fig:STS}(b). Our TB calculations reveal that the LDOS in the VB is dominated by a Te surface state contribution whose location and character is unchanged on altering the thickness, as can be seen in Fig. \ref{fig:Thickness-SM}, and to variation in the alloy fraction.\\\\
The STS measurement of a 5 \si{\nm} thin (Bi$_{0.4}$Sb$_{0.6}$)$_2$Te$_3$ film in Fig. \ref{fig:STS}(c) shows all the important features that can also be recognized in the TB calculations presented in Fig. \ref{fig:STS}(a)-(b). In the TB calculation $E_F$ is locked to the top of the bulk VB around M/6. The recorded STS spectrum however reveals that experimentally, $E_F$ tends to shift to the bottom of the conduction band. In the SM an additional STS measurement is presented, which was performed on another 5 \si{\nm} thin (Bi$_{0.4}$Sb$_{0.6}$)$_2$Te$_3$ film and shows the same weak oscillations in the DOS with a similar peak separation, but with the $E_F$ at a different position, see Fig. \ref{fig:STS-SM}. This discrepancy of the location of the $E_F$ observed when comparing the results from Fig. \ref{fig:STS-SM} with those presented in the main text is more frequently observed when comparing individual spectroscopy results from sample to sample, and even when comparing different terraces on a single sample. This variation can be attributed to either the adsorption process of molecules to the film surface or the presence of defects in the material, such as Te vacancies. To present an estimate of the shift of the $E_F$ that can be induced in the material by Te vacancies, we first make the assumption that $E_F$ is located within the bulk band gap and only crosses the TSS, characterized by a linear dispersion with $v_F =4.7\times10^5$ \si{\metre/\second}. For this dispersion, a shift of 100 \si{\milli\electronvolt} is accompanied by a wave vector change of about $3.2\times10^{8}$ \si{\per\metre}, which corresponds to a 2D carrier density, $n_{2D}$, of $8.3\times10^{11}$ \si{\per\square\cm}. Assuming that every missing Te atom dopes the system with a single electron, this $n_{2D}$ can already be induced when about 0.15\% of the unit cells on the film surface exhibit a single Te vacancy. Therefore we can argue that the shift of about 0.15 \si{\electronvolt} for the comparison in Fig. \ref{fig:STS} is justified.\\\\
The \textit{(dI(V)/dV)/(I(V)/V)} curve, the dashed line in Fig. \ref{fig:STS}(c), shows three distinct peaks that are closely correlated to DOS oscillations found in the TB STS simulation. Looking at the first peak found in the DOS of the STS measurement, at approximately -0.30 \si{\volt}, the TB STS simulation reveals that the origin of this local maximum in DOS originates from the sum of the DOS arising from both the flat bunching bulk states in the $\Gamma-\text{K}$ direction, and the flat Bi/Sb surface state band just below the DP. Since these local maxima occur at an energy very close to $E_D$, it is difficult to point out the exact location of the DP from the signal of the total DOS alone. The second peak in the normalized \textit{dI(V)/dV} curve located at -0.43 \si{\volt} can also be attributed to arise from bunching bulk states. However, for these states the state bunching occurs closer to the K-point along the line $\Gamma-\text{K}$, namely around K/6. The third, broad peak in the normalized \textit{dI(V)/dV} curve stems from the Van Hove singularity of the upper red surface state in the TB calculation in Fig. \ref{fig:STS}(a). These Te-surface states are nearly degenerate in energy over a large part of the BZ in the VC. These may combine to form well localized quasiparticle states in the alloy, which will spread over a range of energies due to the disorder. The predicted van Hove singularity in the VC may thus become broadened in the true alloy. A similar effect will be expected near the minima of the topological surface state with large weight on Bi/Sb. As the disorder is more pronounced in this layer as compared to the Te one, we expect a larger broadening effect here as compared to the Te surface state.\\\\ Since we found that the location of the sub-bands is rather invariant upon changing the film thickness, we are also able to state that the STS measurement matches the results from our ARPES study, presented in Fig. \ref{fig:ARPES}(a). Therefore we can state that the combination of our theoretically predicted electronic structure, using a TB approach, and the spectroscopy experiments provides an adequate and complete picture of the electronic properties of our material system.

\section{Conclusion}
In summary, the TB modelling shows that the bunching of  bulk sub-bands in the alloy (Bi$_{0.4}$Sb$_{0.6}$)$_2$Te$_3$ is likely caused by a Sb/Bi substitution induced change in band order, close to $E_F$ near $\Gamma$, with the TSS still separated from the VB. The fact that the DP lies in close proximity to the bunched bulk sub-bands, which are vulnerable to alloy disorder, makes this material system the perfect platform to investigate the robustness of the TSS against disorder. The close correlations between the TB slab calculations and ARPES measurements allow us to elaborate on the atomic origin of electronic bands near $E_F$, providing a good understanding of the electronic structure of an alloy of the TI (Bi$_{1-x}$Sb$_{x}$)$_2$Te$_3$. By combining the results acquired in the spectroscopy experiments and TB calculations we are able to attribute the oscillatory behavior of the DOS in the STS measurements, performed on 5 \si{nm} (Bi$_{0.4}$Sb$_{0.6}$)$_2$Te$_3$ films, to the flat bunching states with a bulk character and to the Van Hove singularity of the Te surface state in the VB.\\\\
This detailed understanding of the material system will allow for future experiments in the direction of even thinner slabs where a hybridization gap will be induced in the TSS. Devices can be equipped with a gate in order to tune into the hybridization gap that occurs at the Dirac point, hereby enabling access to the topological edge states. The spectroscopy measurements confirmed that the DP is in close proximity to the bulk VB. Therefore, in order to observe signatures of a possible QSH state, for example via magneto-transport measurements, the stoichiometry still shows some room for further optimization. We present Sb$_2$Te$_3$ as another possible candidate to perform future research to the QSH state of matter.

\section{Acknowledgements}
This work is financially supported by the Netherlands Organisation for Scientific Research (NWO) through a VICI grant.

\section*{References}
\bibliography{Bibliography_arXiv}

\begin{thebibliography}{29}
\bibitem [{foo()}]{footnote}%
  \BibitemOpen
  \href@noop {} {\bibinfo {title} {{Which requires a phase factor -1 for the Sb
  5p states relative to the Bi 6p states.}}}\BibitemShut {Stop}%
\end{thebibliography}

\clearpage

\renewcommand{\theequation}{S\arabic{equation}}
\renewcommand{\thefigure}{S\arabic{figure}}
\renewcommand{\bibnumfmt}[1]{[S#1]}
\renewcommand{\citenumfont}[1]{S#1}

\appendix
\onecolumngrid
\begin{center}
	\section*{Supplemental Material for}
  	\textbf{\large Spectroscopic signature of surface states and bunching of bulk subbands in topological insulator (Bi$_{0.4}$Sb$_{0.6}$)$_2$Te$_3$ thin films}\\[.5cm]
  	{ L. Mulder,$^{1, *}$ C. Castenmiller,$^{1, *}$, F.J. Witmans,$^1$ S. Smit,$^2$\\ M.S. Golden,$^2$ H.J.W. Zandvliet,$^1$ P.L. de Boeij$^1$ and A. Brinkman$^1$}\\[.1cm]
  	{\small \itshape ${}^1$MESA+ Institute for Nanotechnology, University of Twente, The Netherlands\\
  	${}^2$Van der Waals-Zeeman Institute, Institute of Physics,  University of Amsterdam, The Netherlands}\\[1cm]
\end{center}
\twocolumngrid

%\section{Appendix}
\setcounter{figure}{0}
\setcounter{equation}{0}
\section*{I. Substrate treatment}
Firstly the SrTiO$_3$ (111) substrates are polished by means of a lens tissue and isopropanol to ensure a clean surface. Following this cleaning step, the SrTiO$_3$ substrates undergo a buffered hydrofluoric acid (BHF) treatment to ensure Ti$^{4+}$ terminated surface. The SrO$_3^{4-}$ surface termination is etched by an ultrasonic immersion of the substrates in demineralized water for 30 \si{\minute}, allowing the formation of SrH compounds. These compounds are subsequently selectively etched by a 30 \si{\second} BHF dip. Next, the substrates are polished again by a lens tissue and isopropanol. To obtain a terraced surface, the chemical treatment is succeeded by a heat treatment in a tube furnace, where the substrates are heated to 950 \si{\degreeCelsius} for 1.5 \si{\hour}, while maintaining an oxygen rich environment by applying an oxygen flow of 150 \si{\milli\litre/\minute} \cite{G.Koster1998APLSM}. The treated substrates from this study typically exhibit terraces with a width of approximately 100 \si{\nm}.\\\\ 
To avoid charging effects during ARPES measurements, as well as to enable a tunneling current to be established on our (Bi$_{0.4}$Sb$_{0.6}$)$_2$Te$_3$ films deposited on insulating SrTiO$_3$, a 5 \si{\nm} W film is sputter deposited on the edge of the substrate by means of a hard mask. These strips are connected to the sample holder, which allows for an electronic connection between the film and the holder. The final substrate treatment step, prior to the film deposition, is an \textit{in-situ} anneal performed for 1 \si{\hour} at 550 \si{\degreeCelsius} in a Te rich environment.

\section*{II. ARPES selection rules to explain asymmetry in EDM}
The ARPES intensity can be linked to the electronic structure that is obtained within our virtual crystal (VC) approximation, by
\begin{equation}
I(\epsilon,\mathbf{k})\propto \sum_{n\sigma} \delta(\epsilon-\epsilon_{n\mathbf{k}}) 
\Big|\mathbf{A}\cdot\mathbf{P}^\sigma_{n\mathbf{k}} \Big|^2.
\label{eq:ArpesIntensity}
\end{equation}
As no spin is detected in the experiment, we need to sum over the spin degrees $\sigma$ of the spinors $\psi^\sigma_{n\mathbf{k}}$ that enter in the transition matrix elements describing the excitation probability from bound to propagating states. These are given by
\begin{equation}
\mathbf{P}^\sigma_{n\mathbf{k}} = \langle e^{i\mathbf{k}\cdot\mathbf{r}- \kappa z}\mid
-i\hbar\nabla \mid \psi^\sigma_{n\mathbf{k}} \rangle.
\label{eq:PMatrixElements}
\end{equation}
This expression reflects the dependence on the polarisation $\mathbf{A}$ of the exciting photons. It also accounts for the conservation of parallel crystal momentum $\mathbf{k}$, while the energy conservation and the energy dependent escape depth of the emitted electron is effectively included through $\kappa(\epsilon,\mathbf{k})$.
For $p$-polarized light at oblique angle of incidence $\theta$, and $\mathbf{k}\parallel \hat{x}$ (along $\Gamma$-K), the transition matrix elements of eq. (\ref{eq:PMatrixElements}) naturally decompose into in-plane $x$-, and out-of-plane $z$-components,
\begin{equation}
\mathbf{A}\cdot\mathbf{P} \propto  P_x \cos\theta +  P_z \sin\theta.
\end{equation}
States that have identical parity for $P_x$ and $P_z$ upon inversion of $\mathbf{k}$ or have exactly one of these elements zero due to symmetry, will show up as symmetric features in the ARPES image, whereas only states with both elements nonvanishing and of opposite parity may show up as asymmetric features, provided symmetry is not restored in the ARPES by multiple degenerate contributions in the summation of eq. (\ref{eq:ArpesIntensity}).
We will now show that (i) the  Bi/Sb topological surface state, which shows up as symmetric, is of the former class, (ii) the Te surface state, which shows up as asymmetric, is of the latter class, and (iii) that the bunched bulk-like states feature is symmetrized by degeneracy.\\
\begin{figure*}
\centering
\includegraphics[width=\textwidth]{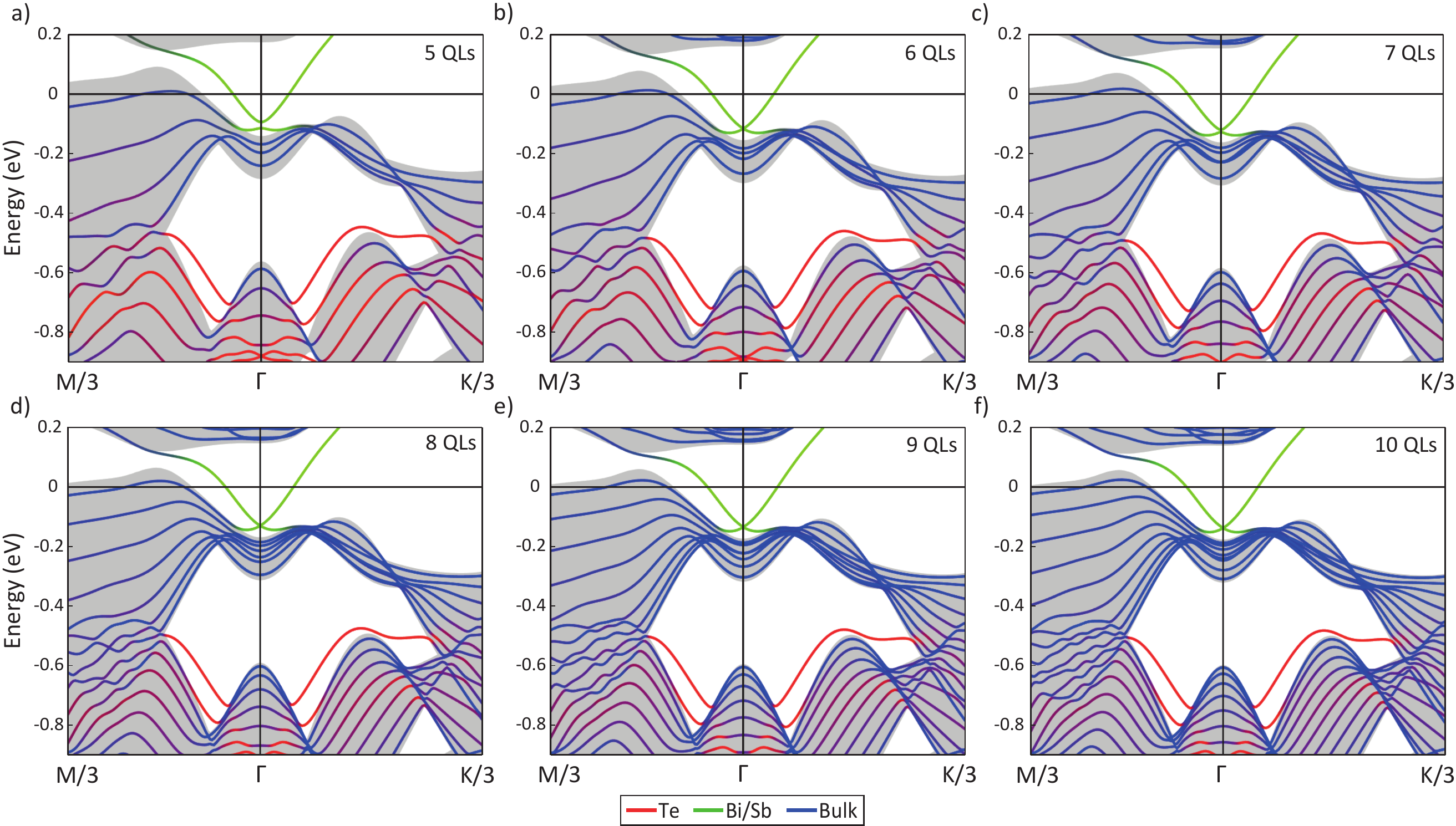}
\caption{\label{fig:Thickness-SM} Calculated band structures for free-standing thin slabs of 5 to 10 QL (Bi$_{0.4}$Sb$_{0.6}$)$_2$Te$_3$. The calculations show that the energy position of the Dirac point in the topological surface states (green lines), as well as the position at which the bulk states bunch (blue lines) and the Te surface state dispersion (red lines) are independent of thickness. This observation allows us to unambiguously compare the ARPES data with the STS data acquired on 10 and 5 \si{\nm} films, respectively.}
\end{figure*} 

\noindent In our VC calculations, we find that the momentum matrix components, 
\begin{equation}
\mathbf{P} = (P_x^\uparrow,P_x^\downarrow)\hat{x}+(P_y^\uparrow, P_y^\downarrow)\hat{y}+(P_z^\uparrow,P_z^\downarrow)\hat{z},
\end{equation}
transform under inversion of the $\mathbf{k}$ labels, into
\begin{equation}
\mathbf{\bar{P}} = e^{i\phi}\Big((-P_y^{\uparrow*},P_y^{\downarrow*})\hat{x}+(P_x^{\uparrow*}, -P_x^{\downarrow*})\hat{y}+
(-P_z^{\downarrow*},P_z^{\uparrow*})\hat{z}\Big).
\end{equation}
It turns out that the contribution of the Bi/Sb topological surface state to the ARPES intensity is well described by a single $\mathbf{P}$-vector,
\begin{equation}
I(\epsilon_{tss},\mathbf{k})\propto  
\Big|\mathbf{A}\cdot\mathbf{P} \Big|^2.
\end{equation}
For small $\mathbf{k}$, we get that $P_y^\uparrow\approx -iP_x^{\uparrow}\in\Re$, $P_y^\downarrow\approx iP_x^{\downarrow}\in\Re$, and $P_z^\uparrow , P_z^{\downarrow}\approx 0$, where the relations become exact for $\mathbf{k} = 0$. As a result, this $\mathbf{P}$ is circularly polarized and oriented in-plane, and thus $\Big|\mathbf{A}\cdot\mathbf{P} \Big|^2\approx\Big|\mathbf{A}\cdot\mathbf{\bar{P}} \Big|^2$ near $\Gamma$. This immediately results in a symmetric appearance of the topological surface state that is visible near $\Gamma$ in the ARPES.\\\\ A similar analysis shows that the same holds for the Te surface state. However, this state is visible at much larger $\mathbf{k}$-vectors, where an asymmetric component in the $p$-polarized ARPES develops due to nonvanishing $P_z$-components away from $\Gamma$, in combination with a notable difference in magnitude for the various spin-up and spin-down components. The $|\mathbf{A}\cdot\mathbf{P}|^2$ at $\mathbf{k}$ and $|\mathbf{A}\cdot\mathbf{\bar{P}}|^2$ at $-\mathbf{k}$ become very different at larger $\mathbf{k}$ due to this out-of-place component, which is not balanced by spin symmetrization. This explains the observed asymmetry.\\\\ Finally we consider the observed symmetric intensity of the nearly degenerate (bunched) bulk-like bands. Their combined contribution to the ARPES intensity near $K/8$ is found to be well described only by a combination of two vectors, being $\mathbf{P}$ and the symmetry related $\mathbf{\bar{P}}$,
\begin{equation}
I(\epsilon_{bbb},\mathbf{k})\propto  
\Big|\mathbf{A}\cdot\mathbf{P} \Big|^2 + \Big|\mathbf{A}\cdot\mathbf{\bar{P}} \Big|^2 .
\end{equation}
Upon inversion of the $\mathbf{k}$-label, the role of $\mathbf{P}$ and $\mathbf{\bar{P}}$ is merely interchanged, resulting in an unaltered intensity. This feature therefore has a $p$-polarized ARPES intensity that is expected to be symmetric under inversion of $\mathbf{k}$.
Finally we remark that in all three cases, we find the magnitudes of the $P$-vectors to be comparable. The observed intensity difference is therefore not explained by this analysis, and should be attributed to other effects.%\newpage

\section*{III. Thickness dependent TB calculations}

Fig. \ref{fig:Thickness-SM} presents \textit{ab-initio} TB calculations for free-standing thin slabs of 5 to 10 QL (Bi$_{0.4}$Sb$_{0.6}$)$_{2}$Te$_{3}$.

\section*{IV. Laser-ARPES}

\begin{figure*}
\centering
\includegraphics[width=0.85\textwidth]{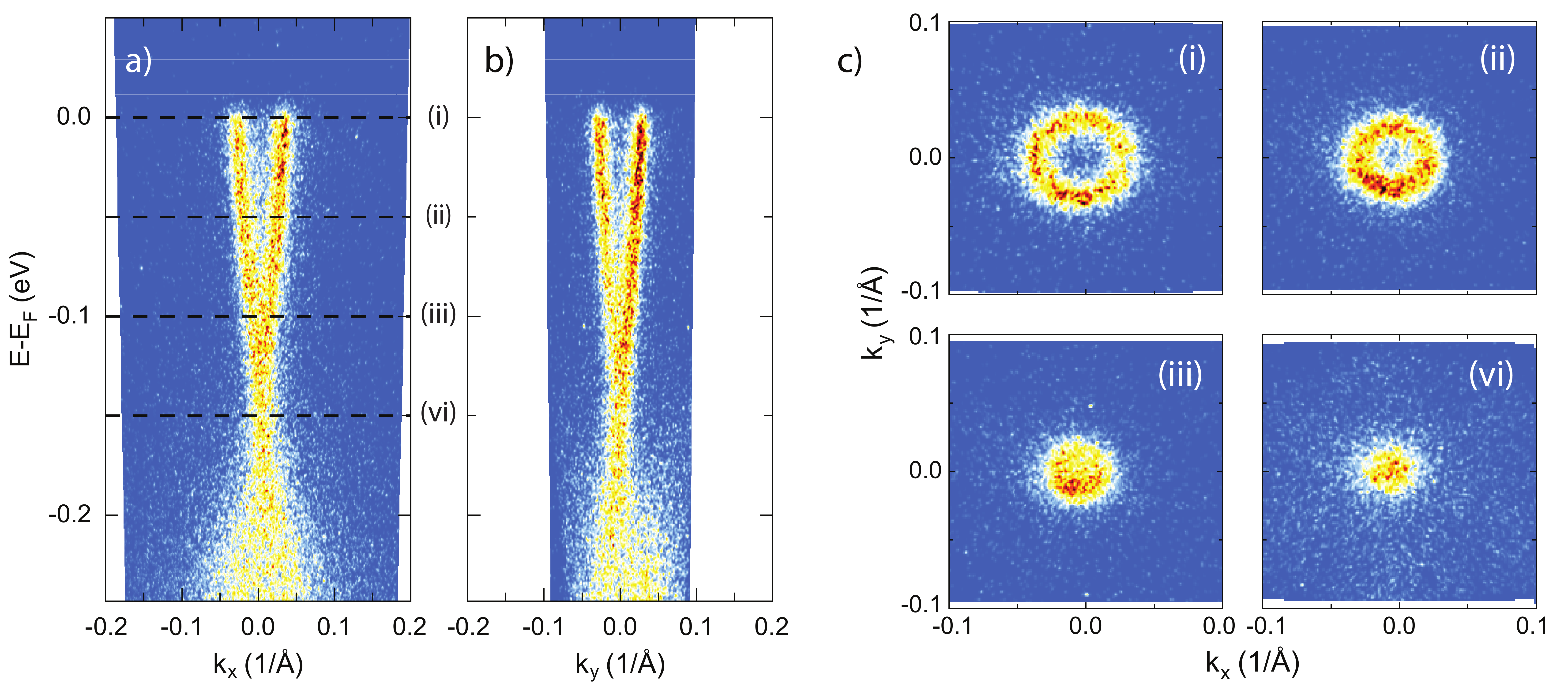}
\caption{\label{fig:ARPES-SM}ARPES spectra of 10 \si{\nm} (Bi$_{0.4}$Sb$_{0.6}$)$_{2}$Te$_{3}$ thin film deposited on SrTiO$_3$, using a photon energy of $h\nu=$ 6.2 \si{\electronvolt}. The EDMs show the clear Dirac cone surface state dispersion in the parallel reciprocal space dimensions, $k_x$ (a) and $k_y$ (b). The FS maps present the constant energy contours at binding energies indicated by the dotted lines in the left EDM (i)-(iv). To enhance the signal to noise ratio, the EDM show the combined signal of a range of 0.03 (1/\si{\angstrom}) in the perpendicular $k_\parallel$ direction, and the FS maps a range of 30 \si{\milli\electronvolt} around $E-E_F=$ 0.0, -0.05, -0.1, and -0.15 \si{\electronvolt}.}
\end{figure*}

Fig. \ref{fig:ARPES-SM} presents Laser-ARPES spectra of a 10 \si{\nm} (Bi$_{0.4}$Sb$_{0.6}$)$_{2}$Te$_{3}$ thin film.

\section*{V. Dirac velocity determination by scanning tunneling spectroscopy}

As shown in ref. \citesupp{Z.Jiao2021APLSM} scanning tunneling spectroscopy can also provide information on the dispersion of the energy bands in momentum space.  By measuring inverse decay length $\kappa(V)$, $k_\parallel$ can be extracted using equations (\ref{tunnel_currents}) and (\ref{eq:kappa}) from the manuscript. Using the relation $E-E_D=e(V-V_D)=\hbar v_F \abs{k}=\hbar v_F k_\parallel$, where $E_D$ refers to the location of the Dirac point, the Fermi velocity, $v_F$, in the vicinity of the Dirac point can be determined. We find a $v_F$ of about 1.5 $\pm$ 0.5$\times 10^5$ \si{\metre/\second}, which is substantially smaller than the $v_F$ extracted from the ARPES spectrum. Bearing in mind that the TB calculations reveal a non-linear dispersion of the topological surface states in the vicinity of the Dirac point one expects a decrease in $v_F$ upon decreasing the energy window near the Dirac point. Therefore, we would like to emphasize that both $v_F$ as extracted from the STS as well as from the ARPES experiments agree very well with the TB calculations. % (3) \ref{tunnel_currents} and (4) \ref{eq:kappa} 
\begin{figure}
\centering
\includegraphics[width=0.48\textwidth]{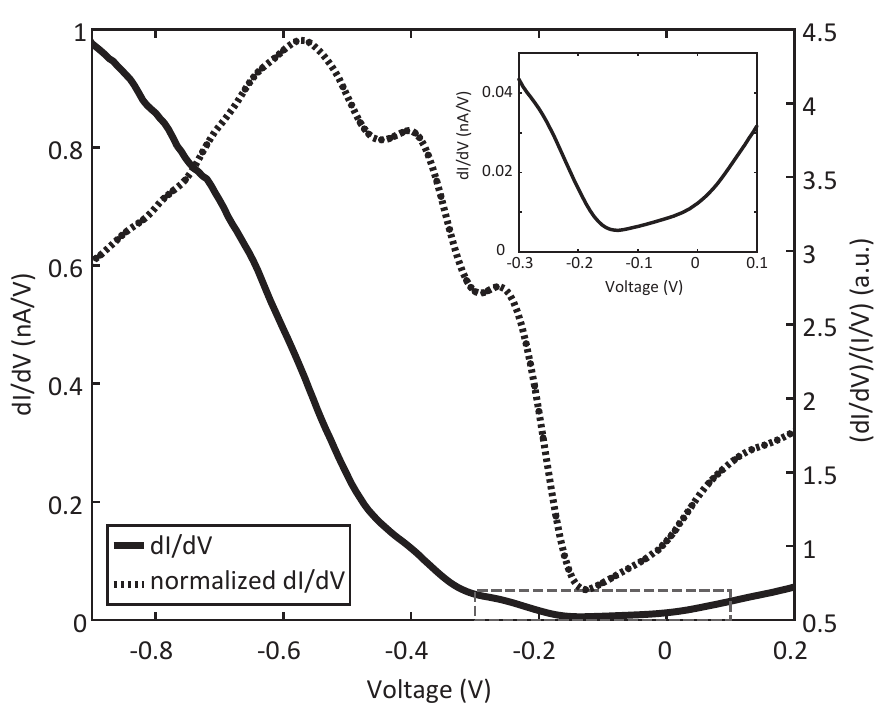}
\caption{\label{fig:STS-SM} STS \textit{dI(V)/dV} and \textit{(dI(V)/dV)/(I/V)} measurements recorded on a 5 \si{\nm} thin (Bi$_{0.4}$Sb$_{0.6}$)$_2$Te$_3$ film (different sample than the measurement from Fig. \ref{fig:STS}). The normalized curve shows peaks at -0.26 \si{\volt}, -0.40 \si{\volt} and -0.57 \si{\volt}. The inset shows a zoom of the \textit{dI(V)/dV} curve, visualized by the dotted box, which shows a minimum in the DOS. The increase in DOS is larger below this minimum than above it due to the presence of the bunched states originating from the (Bi$_{0.4}$Sb$_{0.6}$)$_2$Te$_3$ film.}
\end{figure} 

\section*{VI. Additional scanning tunneling spectroscopy measurement}

Fig. \ref{fig:STS-SM} shows an STS measurement recorded on another 5 \si{\nm} thin (Bi$_{0.4}$Sb$_{0.6}$)$_2$Te$_3$ film than the one presented in Fig. \ref{fig:STS} in the main text. This measurement also shows an oscillatory behavior of the \textit{dI(V)/dV} signal. We can distinguish three clear peaks in the DOS at -0.26 \si{\volt}, -0.40 \si{\volt} and -0.57 \si{\volt}. Although the peak energies are shifted by about 0.04 \si{\volt} with respect to the measurement of Fig. \ref{fig:STS}(c), they can still be explained by the TB slab calculation and STS simulation of Fig. \ref{fig:STS}(a) and (b), respectively. Likewise, the first peak originates from the bunching of states just below the DP, the second peak from the Van Hove singularity of the upper red surface state and the broad third peak from the combined surface states at lower energies. The aforementioned shift of the energies is most likely due to a tiny amount of adsorbates or Te vacancies, which both lead to doping of the thin film. The inset in Fig. \ref{fig:STS-SM} shows a zoom of the \textit{dI(V)/dV} curve around the linear slope observed above the top of the VB. The DOS at -0.13 \si{\volt} is highly reduced with respect to the rest of the bias voltage range. The correlation of the TB calculations with this STS measurement reveals that this linear behavior of \textit{dI(V)/dV} likely originated from the linearly dispersive surface states. The steep slope at bias voltages below the DOS minimum can be ascribed to the presence of the bunched states near the top of the VB.

\section*{References}
%\bibliographysupp{Bibliography_arXiv_SM}

%\bibliography{Bibliography_arXiv}

\end{document}